\documentclass[12pt]{article}
\usepackage{amssymb,amscd}
\usepackage{bm}

\def\thefootnote{\fnsymbol{footnote}}
\newlength{\minitwocolumn}
\catcode`\@=11
\long\def\@makefntext#1{
\protect\noindent \hbox to 3.2pt {\hskip-.9pt  
$^{{\eightrm\@thefnmark}}$\hfil}#1\hfill}               %CAN BE USED 

\def\thefootnote{\fnsymbol{footnote}}
\def\@makefnmark{\hbox to 0pt{$^{\@thefnmark}$\hss}}    %ORIGINAL 
        
\def\ps@myheadings{\let\@mkboth\@gobbletwo
\def\@oddhead{\hbox{}
\rightmark\hfil\eightrm\thepage}   
\def\@oddfoot{}\def\@evenhead{\eightrm\thepage\hfil
\leftmark\hbox{}}\def\@evenfoot{}
\def\sectionmark##1{}\def\subsectionmark##1{}}

\font\eightrm=cmr8

\def\PL{Phys.~Lett. }
\def\PR{Phys.~Rev. }

\newtheorem{thm}{Theorem}[section]

\newtheorem{definition}[thm]{Definition}
\newtheorem{remark}[thm]{Remark}
\newtheorem{proposition}[thm]{Proposition}

\newcommand\gh{{\rm gh}}

\newcommand{\rd}{\overleftarrow{\partial}} 
\newcommand{\ld}{\overrightarrow{\partial}} 

\newcommand{\sbv}[2]{{\{{{#1},{#2}}\}}}

\newcommand{\bracket}[2]{\langle #1\,,#2\rangle}

\def\bx{\mbox{$x$}}
\def\bxi{\mbox{$\xi$}}
\def\bq{\mbox{$q$}}
\def\bp{\mbox{$p$}}
\def\bbx{\mbox{\boldmath $x$}}
\def\bbxi{\mbox{\boldmath $\xi$}}
\def\bbq{\mbox{\boldmath $q$}}
\def\bbp{\mbox{\boldmath $p$}}
\def\bbd{\mbox{\boldmath $d$}}
\def\bbF{\mbox{\boldmath $F$}}
\def\bomega{\mbox{\boldmath $\omega$}}

\newcommand{\calL}{{\cal L}}
\newcommand{\calM}{{\cal M}}
\newcommand{\calN}{{\cal N}}
\newcommand{\calO}{{\cal O}}
\newcommand{\calX}{{\cal X}}

\newcommand{\Q}{{\kern.24em\vrule width.04em height1.4ex%
                 depth-.05ex\kern-.26em\mathsf Q}}
\newcommand{\C}{{\kern.24em\vrule width.04em height1.4ex%
                 depth-.05ex\kern-.26em\mathsf C}}

\newcommand{\Map}{{\rm Map}}
\newcommand{\ev}{{\rm ev}}

\begin{document}

%%%%%%%%%%%%%%%%%%%%%%%%%%%%%%%%%%%%%%%%%%%%%%%%%%%%%%%%%%%%%%%%%%
%%%%%%%%%%%%%%%%%%%%%%%% Title %%%%%%%%%%%%%%%%%%%%%%%%%%%%%%%%%%%
%%%%%%%%%%%%%%%%%%%%%%%%%%%%%%%%%%%%%%%%%%%%%%%%%%%%%%%%%%%%%%%%%%

\baselineskip 0.7cm

\begin{titlepage}
%\today
\begin{flushright}
MISC-2010-01
\end{flushright}

\vskip 1.35cm
\begin{center}
{\Large \bf
QP-Structures
of Degree 3
and 4D Topological Field Theory
}
\vskip 1.2cm
Noriaki IKEDA$^1$%
\footnote{E-mail address:\ ikeda@yukawa.kyoto-u.ac.jp}
and Kyousuke UCHINO$^2$%
\footnote{E-mail address:\ k{\_}uchino@oct.rikadai.jp}
\vskip 0.4cm
{\it 
$^1$
Maskawa Institute for Science and Culture,
%Toshihide Research and Education Center, 
%Maskawa Toshihide Research and Education Center, 
Kyoto Sangyo University, \\
Kyoto 603-8555, Japan \\
%and \\
%%Department of Mathematical Sciences,
%BKC Research Organization of Social Sciences,
%%College of Science and Engineering,
%Ritsumeikan University \\
%Kusatsu, Shiga 525-8577, Japan 
}
\vskip 0.4cm
{\it 
$^2$Department of Mathematics,
Tokyo University of Science\\
Wakamiya 26, Shinjyuku, Tokyo, Japan }

\date{}

\vskip 1.5cm

\begin{abstract}
A BV algebra and a QP-structure of degree $3$ is formulated.
A QP-structure of degree $3$ gives rise to Lie algebroids
up to homotopy and 
its algebraic and geometric structure is analyzed.
A new algebroid is constructed, which derives 
a new topological field theory in $4$ dimensions 
by the AKSZ construction.
\end{abstract}
\end{center}
\end{titlepage}

\renewcommand{\thefootnote}{\alph{footnote}}

\setcounter{page}{2}

%%%%%%%%%%%%%%%%%%%%%%%%%%%%%%%%%%%%%%%%%%%%%%%%%%%%%%%%%%%%%%%%%%
%%%%%%%%%%%%%%%%%%%%%%%% Article %%%%%%%%%%%%%%%%%%%%%%%%%%%%%%%%%
%%%%%%%%%%%%%%%%%%%%%%%%%%%%%%%%%%%%%%%%%%%%%%%%%%%%%%%%%%%%%%%%%%

\rm

%%%%%%%%%%%%%%%%%%%%%%%%%%%%%%%%%%%%%%%%%%%%%%%%%%%%%%%%%%%%%%%%%%%%%
%%%%%%%%%%%%%%%%%%%%%%%%%%%%%%   SEC  1    %%%%%%%%%%%%%%%%%%%%%%%%%%
%%%%%%%%%%%%%%%%%%%%%%%%%%%%%%%%%%%%%%%%%%%%%%%%%%%%%%%%%%%%%%%%%%%%%
\section{Introduction}
\noindent
A BV algebra
%, a QP-manifold 
and a QP-structure 
has been motivated by the structure of 
the Batalin-Vilkovisky formalism of a gauge theory\cite{Bat}
and is its mathematical formulation
\cite{Schwarz:1992nx}.
In case of a topological field theory of Schwarz type, 
a BV formalism has been reformulated to the AKSZ formulation, 
which is 
a clear construction using geometry of a graded manifold 
\cite{Alexandrov:1995kv}\cite{Cattaneo:2001ys}.
Application to higher $n+1$ dimensions has been formulated and 
new topological field theories 
in higher dimensions have been founded
by applying this construction
\cite{Park:2000au}\cite{Severa:2001}\cite{Ikeda:2001fq}.
%\cite{Ikeda:2006wd}.

In $n=1$, 
a classical QP-structure is 
%equivalent to 
a Poisson structure on a manifold $M$ and is also 
a Lie algebroid on $T^*M$ from the explicit construction.
This is equivalent to the construction of a Poisson structure
by the Schouten-Nijenhuis bracket in a classical limit.
The topological field theory in two dimensions constructed 
by the AKSZ formulation \cite{Cattaneo:2001ys}
is the Poisson sigma model
\cite{Ikeda:1993aj}\cite{Schaller:1994es}
and the quantization of this model 
on disc derives the Kontsevich formula of the 
deformation quantization on a Poisson manifold
\cite{Kontsevich:1997vb}\cite{Cattaneo:1999fm}.

In $n=2$, a classical QP-structure is a Courant algebroid
\cite{Courant}\cite{Roy01}.
The topological field theory derived 
in three dimensions is
the Courant sigma model
\cite{Ikeda:2002wh}\cite{Hofman:2002rv}\cite{Roytenberg:2006qz}.

%Though a general theory has been constructed for general $n$,
However structures for higher $n$, more than $2$, 
have not been understood 
enough apart from BF theories.

In this paper, we analyze $n=3$ case.
A QP-structure of degree $3$ leads us to
a new type of algebroid,
which is called a 
\textbf{Lie algebroid up to homotopy}.
%\textbf{4-algebroid}.
The notion of this algebroid
is defined as
a homotopy deformation of a 
Lie algebroid 
satisfying some integrability conditions.
We will prove that a QP-structure of degree $3$
on a N-manifold (nonnegatively graded manifold)
is equivalent to 
a Lie algebroid up to homotopy.
This QP-structure defines
a new natural $4$-dimensional topological field 
theory via the AKSZ construction.

The paper is organized as follows. 
In section 2, a BV algebra and a QP-structure of
degree $3$ are formulated.
In section 3, a QP-structure of degree $3$ is constructed 
and analyzed.
In section 4, examples of QP-structures of degree $3$
are listed.
In section 5, the AKSZ construction of 
a topological field theory in four dimensions
is formulated and examples are listed.
%In section 6, a topological field theory in $4$ dimensions
%is constructed from the QP-structure in section 3 by the AKSZ construction.
%In appendix A, 
%In appendix B, 
\footnote{Very recently, Gr\"utzmann's paper appears which has overlaps
with our paper
\cite{Grutzmann}.}

%%%%%%%%%%%%%%%%%%%%%%%%%%%%%%%%%%%%%%%%%%%%%%%%%%%%%%%%%%%%%%%%%%%%%
%%%%%%%%%%%%%%%%%%%%%%%%%%%%%%   SEC  2    %%%%%%%%%%%%%%%%%%%%%%%%%%
%%%%%%%%%%%%%%%%%%%%%%%%%%%%%%%%%%%%%%%%%%%%%%%%%%%%%%%%%%%%%%%%%%%%%
\section{QP-manifolds and BV Algebras}
%\noindent
\subsection{Classical QP-manifold} 
\begin{definition}
A graded manifold
$\calM$ is by definition a sheaf
of a graded commutative algebra over
an ordinary smooth manifold $M$.
\end{definition}
In the following, we assume the degrees are nonnegative.\\
\indent
The structure sheaf of $\calM$ is locally isomorphic to
a graded commutative algebra $C^{\infty}(U)\otimes S(V)$,
where $U$ is an ordinary local chart of $M$,
$S(V)$ is the polynomial algebra over $V$
and where
$V:=\sum_{i\ge 1}V_{i}$ is a graded vector space
such that the dimension of $V_{i}$ is finite for each $i$.
For example, when $V=V_{1}$, $\calM$ is a vector bundle
whose fiber is $V^{*}_{1}$: the dual space of $V_{1}$.
\begin{definition}
A graded manifold $(\calM,\omega)$ equipped with
a graded symplectic structure $\omega$ of degree $n$
is called a \textbf{P-manifold} of degree $n$.
\end{definition}
In the next section, we will study a concrete 
P-manifold of degree 3.\\
\indent
The structure sheaf $C^{\infty}(\calM)$ of a P-manifold
becomes a graded Poisson algebra.
The Poisson bracket is defined in the usual manner,
\begin{equation}\label{gradedpoisson}
\sbv{F}{G}=(-1)^{|F|+1}\iota_{X_F}\iota_{X_G}\omega,
\end{equation}
where $F,G\in C^{\infty}(\calM)$,
$|F|$ is the degree of $F$
and $X_{F}:=\{F,-\}$ is the Hamiltonian vector field of $F$.
We recall the basic properties of the Poisson bracket,
\begin{eqnarray*}
\sbv{F}{G}&=&-(-1)^{(|F| - n)(|G| - n)} \sbv{G}{F},\\
\sbv{F}{G  H}&=&\sbv{F}{G} H
+ (-1)^{(|F| - n)|G|} G \sbv{F}{H},\\
\{F,\{G,H\}\}&=&\{\{F,G\},H\}+(-1)^{(|F|-n)(|G|-n)}\{G,\{F,H\}\},
\end{eqnarray*}
where $n$ is the degree of the symplectic structure and $F,G,H\in C^{\infty}(\calM)$.
We remark that the degree of the Poisson bracket is $-n$.
\begin{definition}
Let $(\calM,\omega)$ be a P-manifold of degree $n$.
A function $\Theta\in C^{\infty}(\calM)$ of degree $n+1$
is called a \textbf{Q-structure}, if it is a solution of
the \textbf{classical master equation},
\begin{eqnarray}
%\Delta (e^{\frac{i}{\hbar} S}) =0,
\sbv{\Theta}{\Theta}=0.
\label{bvaction}
\end{eqnarray}
The triple $(\calM,\omega,\Theta)$ is called a \textbf{QP-manifold}.
\end{definition}
We define an operator $Q:=\sbv{\Theta}{-}$,
which is called a homological vector field.
From (\ref{bvaction}) we have the cocycle condition,
$$
Q^2=0,
$$
which says that the homological vector field
is a coboundary operator on $C^{\infty}(\calM)$ and 
defines a cohomology called the classical BRST cohomology.
%$H^{\cdot}(C^{\infty}(\calM), Q)$.

%%%%%%%%%%%%%%%%%%%%%%%%%%%%%%%%%%%%%%%%%%%%%%%%%%%%%%%%%%%%%%%%%%%%%%
\subsection{Quantum QP-manifold}
\begin{definition}
%An odd Poisson algebra 
A graded manifold is called a quantum BV-algebra
if it has an odd Laplace operator $\Delta$, 
which is a linear operator on $C^{\infty}(\calM)$
%which is a second order differential operator 
satisfying $\Delta^{2}=0$,
and the graded Poisson bracket is given by
\begin{eqnarray}
\sbv{F}{G}= 
(-1)^{|F|} \Delta(FG) - (-1)^{|F|} \Delta(F){G}-{F}\Delta(G),
\label{Poisson2}
\end{eqnarray}
where $F, G \in C^{\infty}(\calM)$.
%are elements in the Poisson algebra.
\end{definition}
If $n$ is odd, a P-manifold $(\calM,\omega)$ has the odd Poisson bracket.
If an odd P-manifold $(\calM,\omega)$ has a volume form $\rho$,
one can define an odd Laplace operator $\Delta$ 
(See \cite{Khudaverdian:2000zt}):
\begin{eqnarray*}
\Delta F := \frac{1}{2}
(-1)^{|F|} {\rm div}_{\rho} X_F.
%\label{divergence2}
\end{eqnarray*}
%This is an odd Laplace operator satisfying (\ref{Poisson2}).
%The construction is as follows.
Here a divergence ${\rm div}_{\rho}$ 
is a map from a space of vector fields on $\calM$
to $C^{\infty}(\calM)$ and
is defined by
\begin{eqnarray*}
\int_{\calM} {\rm div}_{\rho} X \ F dv = - \int_{\calM} X(F) dv,
\label{divergence}
\end{eqnarray*}
for a vector field $X$ on $\calM$.
%Thus, the Poisson algebra $C^{\infty}(\calM)$ becomes a BV-algebra.
The pair $(\calM, \Delta)$ is called a \textbf{quantum P-structure}.
An odd Laplace operator has degree $-n$.
\begin{definition}
A function $\Theta \in C^{\infty}(\calM)$ with 
the degree $n+1$
is called a \textbf{quantum Q-structure},
if it satisfies a \textbf{quantum master equation}
\begin{eqnarray}
%\Delta (e^{\frac{i}{\hbar} S}) =0,
\Delta (e^{\frac{i}{\hbar}\Theta}) =0,
\label{bvaction2}
\end{eqnarray}
%$\Theta$ is called a {\it quantum BV action},
where $\hbar$ is a formal parameter.
%This is called the quantum master equation.
The triple $(\calM,\Delta,\Theta)$ is called a \textbf{quantum QP-manifold}.
\end{definition}
From the definition of an odd Laplace operator, 
the equation (\ref{bvaction2}) is equivalent to 
\begin{eqnarray}
\sbv{\Theta}{\Theta} - 2 i \hbar \Delta \Theta =0.
%\sbv{S}{S} - 2 i \hbar \Delta S =0
\label{qme}
\end{eqnarray}
%is obtained.
%This is called a quantum master equation.
%
If we take the limit of $\hbar\to 0$ in (\ref{qme}), 
which is called a classical limit,
the classical master equation 
%\begin{eqnarray}
$\sbv{\Theta}{\Theta} =0$
%\label{homological}
%\end{eqnarray}
is derived.
Since $\Delta^2=0$,
$\Delta$ is also a coboundary operator.
This defines a quantum BRST cohomology.
%and defines a complex by the degrees.
%
Let $\calO^{\prime} = \calO e^{\frac{i}{\hbar}\Theta}
\in C^{\infty}(\calM)$ be a cocycle with respect to $\Delta$.
The cocycle condition 
$
\Delta (\calO^{\prime}) 
= \Delta (\calO e^{\frac{i}{\hbar}\Theta}) = 0
$
is equivalent to
\begin{eqnarray}\label{defob}
\sbv{\Theta}{\calO} - i \hbar \Delta \calO = 0.
%\sbv{S}{\calO} - i \hbar \Delta \calO = 0.
\end{eqnarray}
The solutions of (\ref{defob})
%, $\calO$,
are called {\it observables} in physics.
In the classical limit, (\ref{defob}) is 
%\begin{eqnarray*}
$\sbv{\Theta}{\calO} = Q\calO=0$.
%\sbv{S}{\calO} = 0
%\label{clasobs}
%\end{eqnarray*}
${\calO}$ reduces to an element of a classical BRST cohomology.

%%%%%%%%%%%%%%%%%%%%%%%%%%%%%%%%%%%%%%%%%%%%%%%%%%%%%%%%%%%%%%%%%%%%%
%%%%%%%%%%%%%%%%%%%%%%%%%%%%%%%  SEC. 3 %%%%%%%%%%%%%%%%%%%%%%%%%%%%%%
%%%%%%%%%%%%%%%%%%%%%%%%%%%%%%%%%%%%%%%%%%%%%%%%%%%%%%%%%%%%%%%%%%%%%
\section{Structures and homotopy algebroids}

In this section, we construct and analyze a classical QP-structure of
 degree $3$ explicitly.

\subsection{P-structures}

Let $E\to M$ be a vector bundle over an ordinary smooth manifold $M$.
The shifted bundle $E[1]\to M$ is a graded manifold
whose fiber space has the degree $+1$.
We consider the shifted cotangent bundle $\calM:=T^{*}[3]E[1]$.
It is a P-manifold of the degree $3$ over $M$,
$$
T^{*}[3]E[1]\to\calM_{2}\to E[1]\to M,
$$
where $\calM_{2}$ is a certain graded manifold
\footnote{In fact, $\calM_{2}$ is $E[1]\oplus E^*[2]$,
which is derived from the result
in the previous sentence of Remark 3.2.}.
The structure sheaf $C^{\infty}(\calM)$ of $\calM$
is decomposed into the homogeneous subspaces,
$$
C^{\infty}(\calM)=\sum_{i\ge 0}C^{i}(\calM),
$$
where $C^{i}(\calM)$ is the space of functions of degree $i$.
In particular, $C^{0}(\calM)=C^{\infty}(M)$: the algebra of
smooth functions on the base manifold
and $C^{1}(\calM)=\Gamma E^{*}$: the space
of sections of the dual bundle of $E$.\\
\indent
Let us denote by $(x,q,p,\xi)$
a canonical (Darboux) coordinate on $\calM$, where
$x$ is a smooth coordinate on $M$,
$q$ is a fiber coordinate on $E[1]\to M$,
$(\xi,p)$ is the momentum coordinate
on $T^{*}[3]E[1]$ for $(x,q)$.
The degrees of the variables $(x,q,p,\xi)$ are respectively $(0,1,2,3)$.\\
\indent
Two directions of counting the degree of functions
on $T^{*}[3]E[1]$ are introduced. 
Roughly speaking, these are
the fiber direction and the base direction.
\begin{definition}(Bidegree, see also Remark 3.3.3 in \cite{Roy01})
Consider a monomial $\xi^{i}p^{j}q^{k}$ on a local chart
$(U;x,q,p,\xi)$ of $\calM$, of which the total degree is $3i+2j+k$.
The \textbf{bidegree} of the monomial is, by definition,
$(2(i+j),i+k)$.
\end{definition}
This definition is invariant under the natural coordinate transformation,
\begin{eqnarray*}
x^{\prime}_{i}&=&x^{\prime}_{i}(x_{1},x_{2},...,x_{dim(M)}),\\
q^{\prime}_{i}&=&\sum_{j}t_{ij}q_{j},\\
p^{\prime}_{i}&=&\sum_{j}t^{-1}_{ij}p_{j},\\
\xi^{\prime}_{i}&=&\sum_{j}\frac{\partial x_{j}}{\partial x^{\prime}_{i}}\xi_{j}
+\sum_{jkl}
(\frac{\partial t^{-1}_{jl}}{\partial x^{\prime}_{i}}t_{lk}
+\frac{\partial t_{kl}}{\partial x^{\prime}_{i}}t^{-1}_{lj})
p_{j}q_{k},
\end{eqnarray*}
where $t$ is a transition function. Since $T^{*}[3]E[1]$ is covered
by the natural coordinates,
the bidegree is globally well-defined
(See also Remark \ref{shiftremark} below.)\\
\indent
The space $C^{n}(\calM)$ is uniquely decomposed into
the homogeneous subspaces with respect to the bidegree,
$$
C^{n}(\calM)=\sum_{2i+j=n}C^{2i,j}(\calM).
$$
Since $C^{2,0}(\calM)=\Gamma E$ and $C^{0,2}(\calM)=\Gamma\wedge^{2}E^{*}$,
we have
$$
C^{2}(\calM)=\Gamma E\oplus\Gamma\wedge^{2}E^{*}.
$$
\begin{remark}\label{shiftremark}
The P-manifold $T^{*}[3]E[1]$ is regarded as
a shifted manifold of $T^{*}[2]E[1]$.
%Hence t
The structure sheaf is also a shifted 
sheaf of the one on $T^{*}[2]E[1]$.
In particular, the space $C^{2i,j}$ is 
the shifted space of $C^{i,j}$ on $T^{*}[2]E[1]$.
\end{remark}
For the canonical coordinate on $\calM$,
the symplectic structure has the following form:
$$
\omega = \delta \bx^i \delta \bxi_i + \delta \bq^a \delta \bp_a,
$$
and the associated Poisson bracket
has the following expression:
\begin{eqnarray*}
%\label{BVbracket}
\sbv{F}{G} &=& 
F \frac{\rd}{\partial \bx^i} 
\frac{\ld }{\partial \bxi_{i}} G
- 
F \frac{\rd }{\partial \bxi_{i}} 
\frac{\ld }{\partial \bx^i} G
+ 
F  \frac{\rd}{\partial \bq^{a}} 
\frac{\ld }{\partial \bp_{a}} G
- 
F \frac{\rd }{\partial \bp_{a}} 
\frac{\ld }{\partial \bq^{a}} G,
\end{eqnarray*}
where $F,G\in C^{\infty}(\calM)$
and $\frac{\overrightarrow{\scriptstyle{\partial}}}{\partial \phi}$ and 
$\frac{\overleftarrow{\scriptstyle{\partial}}}{\partial \phi}$
are the right and left differentiations, respectively.
Note that the degree of the symplectic structure is $+3$
and the one of the Poisson bracket is $-3$.
The bidegree of the Poisson bracket is $(-2,-1)$, that is,
$$
\{(2i,j),(2k,l)\}=(2(i+k)-2,j+l-1),
$$
where $(2i,j)$... are functions with the bidgree $(2i,j)$.

%%%%%%%%%%%%%%%%%%%%%%%
\subsection{Q-structures}
%%%%%%%%%%%%%%%%%%%%%%%%%
We consider a (classical) Q-structure, $\Theta$, on the P-manifold.
It is required that $\Theta$ has degree $4$.
That is, $\Theta\in C^{4}(\calM)$.
Because $C^{4}(\calM)=C^{4,0}(\calM)\oplus C^{2,2}(\calM)\oplus C^{0,4}(\calM)$,
the Q-structure is uniquely decomposed into
$$
\Theta=\theta_{2}+\theta_{13}+\theta_{4},
$$
where the bidegrees of the substructures are
$(4,0)$, $(2,2)$ and $(0,4)$, respectively.
In the canonical coordinate, $\Theta$ is the following polynomial:
\begin{eqnarray}\label{deftheta}
\Theta = f{}_1{}^i{}_{a} (\bx) \bxi_i \bq^a 
+\frac{1}{2} f_2{}^{ab}(\bx) \bp_a \bp_b
+\frac{1}{2} f_3{}^a{}_{bc}(\bx) \bp_a \bq^b \bq^c
+\frac{1}{4!} f_4{}_{abcd}(\bx) \bq^a \bq^b \bq^c \bq^d,
\end{eqnarray}
and the substructures are 
\begin{eqnarray*}
\theta_{2}&=&\frac{1}{2} f_2{}^{ab}(\bx) \bp_a \bp_b,\\
\theta_{13}&=&
f{}_1{}^i{}_{a} (\bx) \bxi_i \bq^a+\frac{1}{2} f_3{}^a{}_{bc}(\bx) \bp_a \bq^b \bq^c,\\
\theta_{4}&=&\frac{1}{4!} f_4{}_{abcd}(\bx) \bq^a \bq^b \bq^c \bq^d,
\end{eqnarray*}
%respectively, 
where $f_{1}$-$f_{4}$ are structure functions on $M$.
By  counting the bidegree, one can easily prove that
the classical master equation $\sbv{\Theta}{\Theta} = 0$
is equivalent to the following three identities:
\begin{eqnarray}
\label{tc1}
\{\theta_{13},\theta_{2}\}&=&0,\\
\label{tc2}
\frac{1}{2}\{\theta_{13},\theta_{13}\}
+\{\theta_{2},\theta_{4}\}&=&0,\\
\label{tc3}
\{\theta_{13},\theta_{4}\}&=&0.
\end{eqnarray}
The conditions (\ref{tc1}), (\ref{tc2}) and (\ref{tc3})
are equivalent to
\begin{eqnarray}
\label{fc1}
&&f{}_1{}^i{}_{b} f_2{}^{ba} = 0,\\
\label{fc2}
&&
f{}_1{}^k{}_{c} \frac{\partial f_2{}^{ab}}{\partial x^k} 
+ f_2{}^{da} f_3{}^b{}_{cd} + f_2{}^{db} f_3{}^a{}_{cd} = 0,\\
\label{fc3}
&&
f{}_1{}^k{}_{b} \frac{\partial f{}_1{}^i{}_{a}}{\partial x^k} 
- f{}_1{}^k{}_{a} \frac{\partial f{}_1{}^i{}_{b}}{\partial x^k} 
+ f{}_1{}^i{}_{c} f_3{}^c{}_{ab} = 0,\\
\label{fc4}
&& 
f{}_1{}^k{}_{[d} \frac{\partial f_3{}^a{}_{bc]}}{\partial x^k} 
+ f_2{}^{ae} f_4{}_{bcde}
- f_3{}^a{}_{e[b} f_3{}^e{}_{cd]} = 0,\\
\label{fc5}
&& 
f{}_1{}^k{}_{[a} \frac{\partial f_4{}_{bcde]}}{\partial x^k} 
+ f_3{}^f{}_{[ab} f_4{}_{cde]f} =0,
\end{eqnarray}
where $[b \ c \ d \ \cdots]$ is a skewsymmetrization
with respect to indices $b, c, d, \cdots$, etc.
%%%%%%%%%%%%%%%%%%%%%%%%%%%%%%%%%
\subsection{Lie algebroid up to homotopy}
%%%%%%%%%%%%%%%%%%%%%%%%%%%%%%%%%
In this section we study an algebraic structure
associated with the QP-structure in 3.1 and 3.2.
\begin{definition}
Let $Q=\theta_{2}+\theta_{13}+\theta_{4}$ be a $Q$-structure
on $T^{*}[3]E[1]$, where $(\theta_{2},\theta_{13},\theta_{4})$
is the unique decomposition of $\Theta$.
We call the quadruple $(E; \theta_{2},\theta_{13},\theta_{4})$
a \textbf{Lie algebroid up to homotopy},
%or shortly, 
in shorthand, Lie algebroid u.t.h.
\end{definition}
We should study the algebraic properties of the Lie algebroid up to homotopy.
Let us define a bracket product by
\begin{equation}\label{braee}
[e_{1},e_{2}]:=\{\{\theta_{13},e_{1}\},e_{2}\},
\end{equation}
where $e_{1},e_{2}\in\Gamma E$.
By the bidegree counting, 
%one can easily check that
$\Gamma E$ is closed under this bracket.
The bracket is not necessarily a Lie bracket,
but it is still skewsymmetric:
\begin{eqnarray*}
[e_{1},e_{2}]&=&\{\{\theta_{13},e_{1}\},e_{2}\},\\
&=&\{\theta_{13},\{e_{1},e_{2}\}\}+\{e_{1},\{\theta_{13},e_{2}\}\},\\
&=&-\{\{\theta_{13},e_{2}\},e_{1}\}=-[e_{2},e_{1}],
\end{eqnarray*}
where $\{e_{1},e_{2}\}=0$ is used.
A bundle map $\rho:E\to TM$ which is called an anchor map
is defined by the following identity:
% below,
$$
\rho(e)(f):=\{\{\theta_{13},e\},f\},
$$
where $f\in C^{\infty}(M)$.
The bracket and the anchor map 
satisfy the algebroid conditions (A0) and (A1) below:
\begin{description}
\item[(A0)]
$\rho[e_{1},e_{2}]=[\rho(e_{1}),\rho(e_{2})]$,
\item[(A1)]
$[e_{1},fe_{2}]=f[e_{1},e_{2}]+\rho(e_{1})(f)e_{2}$,
\end{description}
where the bracket $[\rho(e_{1}),\rho(e_{2})]$
is the usual Lie bracket on $\Gamma TM$.
The bracket (\ref{braee}) does not satisfy the Jacobi identity in general.
So we should study its Jacobi anomaly, which characterizes
the algebraic structure of the Lie algebroid u.t.h.
The structures $\theta_{13}$, $\theta_{2}$ and $\theta_{4}$ define
the three operations:
\begin{itemize}
\item $\delta(-):=\{\theta_{13},-\}$;
a de Rham type derivation on $\Gamma\wedge^{\cdot}E^{*}$,
\item $(\alpha_{1},\alpha_{2}):=\{\{\theta_{2},\alpha_{1}\},\alpha_{2}\}$;
a symmetric pairing on $E^{*}$, where $\alpha_{1},\alpha_{2}\in\Gamma E^{*}$,
\item $\Omega(e_{1},e_{2},e_{3},e_{4}):=
\{\{\{\{\{\theta_{4},e_{1}\},e_{2}\},e_{3}\},e_{4}\}$;
a 4-form on $E$.
\end{itemize}
Remark that $\delta\delta\neq 0$ in general.
Because the degree of the pairing is $-2$,
it is $C^{\infty}(M)$-valued.
The pairing induces a symmetric
bundle map $\partial:E^{*}\to E$
which is defined by the equation,
$(\alpha_{1},\alpha_{2})=\bracket{\partial \alpha_{1}}{\alpha_{2}}$,
where $\bracket{-}{-}$ is the canonical pairing
of the duality of $E$ and $E^{*}$.
Since $\bracket{\alpha}{e}=\{\alpha,e\}$, we have
$$
\partial\alpha=-\{\theta_{2},\alpha\}.
$$
By direct computation, we obtain
$$
\frac{1}{2}
\{\{\{\{\theta_{13},\theta_{13}\},e_{1}\},e_{2}\},e_{3}\}
=[[e_{1},e_{2}],e_{3}]+({\rm cyclic \ permutations}),
$$
and
$$
\{\{\{\{\theta_{2},\theta_{4}\},e_{1}\},e_{2}\},e_{3}\}
=-\partial\Omega(e_{1},e_{2},e_{3}).
$$
From Eq.~(\ref{tc2}), we get an explicit formula of the Jacobi anomaly,
\begin{description}
\item[(A2)]
$[[e_{1},e_{2}],e_{3}]+({\rm cyclic \ permutations})
= \partial\Omega(e_{1},e_{2},e_{3})$.
\end{description}
In a similar way, we obtain the following identities:
\begin{description}
\item[(A3)] $\rho\partial=0$,
\item[(A4)] $\rho(e)(\alpha_{1},\alpha_{2})=(\mathcal{L}_{e}\alpha_{1},\alpha_{2})
+(\alpha_{1},\mathcal{L}_{e}\alpha_{2})$,
\item[(A5)] $\delta\Omega=0$,
\end{description}
where $\mathcal{L}_{e}(-):=\{\{\theta_{13},e\},-\}$
is the Lie type derivation which acts on $E^{*}$.
Axioms (A3) and (A4) are induced from Eq.~(\ref{tc1})
and (A5) is from Eq.~(\ref{tc3}).\\
\indent
The fundamental relations (\ref{fc1})--(\ref{fc5})
correspond to Axioms (A1)--(A5)\footnote{
Actually, the axiom (A0) depends on (A1) and (A2).
}.
Thus, the notion of the Lie algebroid up to homotopy
is characterized by the algebraic properties (A1)--(A5).
One concludes that
\medskip\\
\noindent
{\em
The classical algebra associated with the QP-manifold $(T^{*}[3]E[1],\Theta)$
is the space of sections of the vector bundle $E$ with the operations
$([\cdot,\cdot],\rho,\partial,\Omega)$
satisfying (A1)--(A5).
}
\medskip\\
\indent
In the next section, we will study some special examples
of Lie algebroid u.t.h.s.
\begin{remark}
\normalfont
If the pairing is nondegenerate,
then the bundle map $\partial$ is bijective
and then from (A3) we have $\rho=0$.
\end{remark}
\begin{remark}\label{CDB}
\normalfont
(Higher Courant-Dorfman brackets)
We define a bracket on $C^{\infty}(\calM)$ by
$$
[-,-]_{CD}:=\{\{\Theta,-\},-\},
$$
which is called a Courant-Dorfman (CD) bracket.
It is well-known that $[,]_{CD}$ is a Loday bracket (\cite{Kos}).
Since the degree of the CD-bracket is $-2$,
the total space of degree $i\le 2$,
$$
C^{2}(\calM)\oplus C^{1}(\calM)\oplus C^{0}(M)
$$
is closed under the CD-bracket, in particular,
the top space $C^{2}(\calM)=\Gamma(E\oplus\wedge^{2}E^{*})$
is a subalgebra.
If $\theta_{2}=0$, the CD-bracket on $E\oplus\wedge^{2}E^{*}$
has the following form,
$$
[e_{1}+\beta_{1},e_{2}+\beta_{2}]_{CD}=
[e_{1},e_{2}]+\mathcal{L}_{e_{1}}\beta_{2}-i_{e_{2}}\delta\beta_{1}+\Omega(e_{1},e_{2}),
$$
where $\beta_{1},\beta_{2}\in\Gamma\wedge^{2}E^{*}$.
This CD-bracket is regarded as a higher analogue of
Courant-Dofman's original bracket (cf. \cite{Courant}).
We refer the reader to 
Hagiwara \cite{Hagiwara} and Sheng \cite{YS}
for the detailed study of the higher CD-brackets.
\end{remark}

%%%%%%%%%%%%%%%%%%%%%%%%%%%%%%%%%%%%%%%%%%%%%%%%%%%%%%%%%%%%%%%%%%%%%
%%%%%%%%%%%%%%%%%%%%%%%%%%%%%%%  SEC. 4 %%%%%%%%%%%%%%%%%%%%%%%%%%%%%%
%%%%%%%%%%%%%%%%%%%%%%%%%%%%%%%%%%%%%%%%%%%%%%%%%%%%%%%%%%%%%%%%%%%%%
\section{Examples and twisting transformations}

\subsection{The cases of $\theta_{2}=\theta_{4}=0$}

In this case, the bracket (\ref{braee}) satisfies
(A0), (A1) and the Jacobi identity.
Therefore, the bundle $E\to M$ becomes a Lie algebroid:
\begin{definition}
(\cite{Mackenzie})
A Lie algebroid over a manifold $M$ is a vector bundle
$E \rightarrow M$ with a Lie algebra structure on the 
space of the sections $\Gamma(E)$ defined by the 
bracket $[e_1, e_2]$ for $e_1, e_2 \in \Gamma(E)$
and an anchor map
$\rho: E \rightarrow TM$ satisfying (A0) and (A1) above.
\end{definition}
%the following properties:
%\begin{eqnarray}
%\label{liealgdef1}\rho[e_1, e_2]&=&[\rho(e_1), \rho(e_2)]
%\label{liealgdef2}[e_1, f e_2]&=&f [e_1, e_2] + (\rho(e_1) f) e_2,
%\end{eqnarray}
%where $e_1$ and $e_2$ are sections of $E$ and 
%$f \in C^{\infty}(M)$.\\
\indent
We take $\{ e_a \}$ as a local basis of $\Gamma E$ and
let a local expression of an anchor map
be $\rho(e_a) = f^i{}_{1a}(x) \frac{\partial}{\partial x^i}$
and a Lie bracket be
$[e_b, e_c] = f_3{}^a{}_{bc}(x) e_a$.
The Q-structure $\Theta$ associated with
the Lie algebroid $E$
is defined as a function on $T^{*}[3]E[1]$,
$$
\Theta:=\theta_{13}:=f{}_1{}^i{}_{a} (\bx) \bxi_i \bq^a
+\frac{1}{2} f_3{}^a{}_{bc}(\bx) \bp_a \bq^b \bq^c,
$$
which is globally well-defined.
Conversely, if we consider $\Theta:=\theta_{13}$,
the classical master equation
induces the Lie algebroid structure on $E$.
%We note that the equations 
%(\ref{liealgdef1}) and (\ref{liealgdef2}) are the 
%same as 
%the conditions (A0) and (A1).
\medskip\\
\indent
Let us consider the case that the bundle is 
a vector space on a point.
A Lie algebroid over
a point $\mathfrak{g}\to\{pt\}$ is
a Lie algebra $\mathfrak{g}$.
The P-manifold over $\mathfrak{g}\to\{pt\}$
is isomorphic to $\mathfrak{g}^*[2]\oplus\mathfrak{g}[1]$
and the structure sheaf is
the polynomial algebra over $\mathfrak{g}[2]\oplus\mathfrak{g}^{*}[1]$,
$$
C^{\infty}(\calM)=S(\mathfrak{g})\otimes \bigwedge^{\cdot}\mathfrak{g}^{*}.
$$
The bidegree is defined by the natural manner,
$$
C^{2i,j}(\calM)=S^{i}(\mathfrak{g})\otimes \bigwedge^{j}\mathfrak{g}^{*}.
$$
The Q-structure associated with the Lie bracket on $\mathfrak{g}$ is
\begin{eqnarray}
\theta_{13}=\frac{1}{2} f^a{}_{bc} p_a q^b q^c
\cong\frac{1}{2} f^a{}_{bc} p_a \otimes(q^b \wedge q^c),
\label{liealgebra}
\end{eqnarray}
where $p_{\cdot}\in\mathfrak{g}$, $q_{\cdot}\in\mathfrak{g}^{*}$
and $f^{a}{}_{bc}$ is the structure constant of the Lie algebra.

\subsection{The cases of $\theta_{2}\neq 0$ and $\theta_{4}=0$}

In this case, the bracket induced by $\theta_{13}$
still satisfies the Jacobi identity.
\medskip\\
\indent
We assume that $\frak{g}$ is semi-simple.
Then the dual space $\frak{g}^{*}$ has a metric,
$(\cdot,\cdot)_{K^{-1}}$, which is
the inverse of the Killing form on $\frak{g}$.
%Since the Killing form is invariant,
%the metric also satisfies an invariant condition
%with respect to the coadjoint action,
The metric inherits the following invariant condition from
the Killing form:
\begin{equation}\label{la3}
(\mathcal{L}_{p}q_{1},q_{2})_{K^{-1}}
+(q_{1},\mathcal{L}_{p}q_{2})_{K^{-1}}=0,
\end{equation}
where $\mathcal{L}_{p}(-)$ is the canonical
coadjoint action of $\mathfrak{g}$ to $\mathfrak{g}^{*}$.
%We notice that 
Eq.~(\ref{la3}) is a linear version of (A4).
Thus, we obtain a Q-structure,
\begin{eqnarray}
\Theta:=
k^{ab}\bp_a \bp_b
+\frac{1}{2} f{}^a{}_{bc} \bp_a \bq^b \bq^c,
\label{killingQ}
\end{eqnarray}
where $k^{ab}\bp_a \bp_b:=(\cdot,\cdot)_{K^{-1}}$.

%%%%%%%%%%%%%%%%%%%%%%%%%%%%%%%%%%%%
\subsection{Non Lie algebra example}
%%%%%%%%%%%%%%%%%%%%%%%%%%%%%%%%%%%%%
We consider the cases that the Jacobi identity is broken.
Let $(\mathfrak{g},[\cdot,\cdot],(\cdot,\cdot)_{K})$
be a vector space (not necessarily Lie algebra)
equipped with a skewsymmetric bracket $[\cdot,\cdot]$
and an invariant metric $(\cdot,\cdot)_{K}$.
The metric induces a bijection
$K:\mathfrak{g}\to\mathfrak{g}^{*}$
which is defined by the identity,
$$
(p_{1},p_{2})_{K}=\bracket{Kp_{1}}{p_{2}}.
$$
We define a map from
$\mathfrak{g}^{*}$ to $\mathfrak{g}$ by $\partial:=K^{-1}$
and define a 4-form by,
$$
\Omega(p_{1},p_{2},p_{3},p_{4}):=\Big([[p_{1},p_{2}],p_{3}]+
{\rm cyclic \ permutations},p_{4}\Big)_{K}.
$$
\begin{remark}
The 4-form above is considered to be a higher analogue
of the Cartan 3-form $([p_{1},p_{2}],p_{3})_{K}$.
\end{remark}
Axioms (A0)--(A4) obviously hold on $\frak{g}$.
We check (A5). It suffices to show (\ref{tc3}).
Let us denote by $\{-,p_{1},p_{2},...,p_{n}\}$
the n-fold bracket $\{...\{\{-,p_{1}\},p_{2}\},...,p_{n}\}$.
We already have (\ref{tc1}) and (\ref{tc2}).
From $\{\theta_{13},\{\theta_{13},\theta_{13}\}\}=0$
and (\ref{tc2}), we have $\{\theta_{13},\{\theta_{2},\theta_{4}\}\}=0$.
Since $\{\theta_{13},\theta_{2}\}=0$,
this is equal to
$\{\theta_{2},\{\theta_{3},\theta_{4}\}\}=0$
up to sign.
This gives
$\{\{\theta_{2},\{\theta_{3},\theta_{4}\}\},p_{1},...,p_{5}\}=0$
for any $p_{1},...,p_{5}$.
From $\{\theta_{2},p\}=0$, we have
$$
\{\theta_{2},\{\{\theta_{3},\theta_{4}\},p_{1},...,p_{5}\}\}=0.
$$
Since $K^{-1}=-\{\theta_{2},-\}$ is bijective, we get
$$
\{\{\theta_{3},\theta_{4}\},p_{1},...,p_{5}\}=0,
$$
which yields the desired relation $\{\theta_{3},\theta_{4}\}=0$.
\begin{proposition}
The triple $\big(\mathfrak{g},\partial,\Omega\big)$
is a Lie algebra(oid) up to homotopy.
\end{proposition}
%%%%%%%%%%%%%%%%%%%%%%%%%%%%%%%%%
\subsection{Twisting by $3$-form and the cases of $\theta_{2}=0$ and $\theta_{4}\neq 0$}
%%%%%%%%%%%%%%%%%%%%%%%%%%%%%%%%
We introduce the notion of twisting transformation by $3$-form
before studying the cases of $\theta_{2}=0$.
Given a Q-structure $\Theta$ and a 3-form $\phi\in C^{0,3}(\calM)$,
there exists the second Q-structure which is defined by
the canonical transformation,
\begin{equation}\label{defgauge}
\Theta^{\phi}:=\exp(X_{\phi})(\Theta),
\end{equation}
where $X_{\phi}:=\{\phi,-\}$ is the Hamiltonian vector field of $\phi$.
The transformation (\ref{defgauge}) is called a \textbf{twisting by 3-form},
or simply twisting.
By a direct computation, we obtain
\begin{eqnarray*}
\label{g1}\theta^{\phi}_{2}&=&\theta_{2},\\
\label{g2}\theta^{\phi}_{13}&=&\theta_{13}-\{\theta_{2},\phi\},\\
\label{g3}\theta^{\phi}_{4}&=&\theta_{4}-\{\theta_{13},\phi\}+\frac{1}{2}\{\{\theta_{2},\phi\},\phi\},
\end{eqnarray*}
where $\Theta^{\phi}=\theta^{\phi}_{2}+\theta^{\phi}_{13}+\theta^{\phi}_{4}$
and $X^{i\ge 3}_{\phi}(\Theta)=0$.
The twisting by 3-form defines an equivalence relation on the Q-structures.
\medskip\\
\indent
We notice that $\theta_{2}$ is an invariant for the twisting.
If $\theta_{2}=0$, then
$\theta_{13}$ is an invariant and
$$
\theta^{\phi}_{4}=\theta_{4}-\delta\phi,
$$
where $\delta\phi=\{\theta_{13},\phi\}$.
This leads us
\begin{proposition}
The class of Q-structures which have no $\theta_{2}$
is classified into $H^{4}_{dR}(\bigwedge^{\cdot}E^{*},\delta)$
by the twisting by 3-form.
\end{proposition}

%%%%%%%%%%%%%%%%%%%%%%%%%%%%%%%%%%%%%%%%%%%%%%%%%%%%%%%%%%%%%%%%%%%%%
%%%%%%%%%%%%%%%%%%%%%%%%%%%%%%%  SEC. 5 %%%%%%%%%%%%%%%%%%%%%%%%%%%%%%
%%%%%%%%%%%%%%%%%%%%%%%%%%%%%%%%%%%%%%%%%%%%%%%%%%%%%%%%%%%%%%%%%%%%%
\section{AKSZ Construction of Topological Field Theory 
in $4$ Dimensions
}
%%%%%%%%%%%%%%%%%%%%%%%%%%%
\subsection{General Theory}
%%%%%%%%%%%%%%%%%%%%%%%%%%%

%\subsection{QP-Structure on $\Map(\calX, \calM)$}
\noindent
In this section, we consider the AKSZ construction of 
a topological field theory in $4$ dimensions.

For a graded manifold $\calN$,  
let $\calN|_0$ be the degree zero part.
%of the graded manifold $\calX$.
%$\calM|_0$ is that of the graded manifold $\calM$ respectively.

%
Let $X$ be a manifold in $4$ dimensions
%$X$ is called a worldsheet if $n=2$, 
%or a worldvolume if $n >2$.
and $M$ be a manifold in $d$ dimensions.
%$M$ is a called a target space.
Let $(\calX, D)$ be a differential graded (dg) manifold 
$\calX$
with a $D$-invariant nondegenerate measure $\mu$, 
such that $\calX|_0 = X$, where
$D$ is a differential on $\calX$.
($\calM, \omega, \Theta$)
is a QP-manifold of degree $3$ and
$\calM|_0 = M$.
A degree $\deg (-)$ on $\calX$ is called the {\it form degree} and 
a degree $\gh (-)$ on $\calM$ is called the {\it ghost number}
\footnote{The ghost number $\gh (-)$ is the degree $|-|$
on $\calM$ in section 2.}.
Let 
$\Map(\calX, \calM)$ be
%$\bbx:X \longrightarrow M \in 
a space of smooth maps from $\calX$ to $\calM$.
$|-| = \deg (-) + \gh (-)$ is the degree
on $\Map(\calX, \calM)$ and called the {\it total degree}.
A QP-structure on $\Map(\calX, \calM)$
is constructed from the above data.

Since ${\rm Diff}(\calX)\times {\rm Diff}(\calM)$ 
naturally acts on $\Map(\calX, \calM)$,
$D$ and $Q$ induce homological vector fields 
on $\Map(\calX, \calM)$, 
$\hat{D}$ and $\check{Q}$. 
%
%We consider $\calX = T[1]X$
%In order to construct a QP-structure on $\Map(\calX, \calM)$,

Two maps are introduced.
An {\it evaluation map} 
${\rm ev}: \calX \times \calM^{\calX} \longrightarrow \calM$ 
is defined as
\begin{eqnarray*}
{\rm ev}:(z, \Phi) \longmapsto \Phi(z),
\end{eqnarray*}
where 
$z \in \calX$ and $\Phi \in \calM^{\calX}$.

A {\it chain map} $\mu_*: \Omega^{\bullet}(\calX \times \calM) 
\longrightarrow \Omega^{\bullet}(\calM)$ is defined as 
$\mu_* F = \int_{\calX} \mu F$
where $F \in \Omega^{\bullet}(\calX \times \calM)$ 
and 
$\int_{\calX} \mu$  is an integration on $\calX$
by the $D$-invariant measure $\mu$.
It is an usual integral for the even degree parts
and
the Berezin integral for the 
odd degree parts. 
%such that
%$\int_{\rm fiber} d \theta^{\mu} \ \theta^{\nu} = \delta^{\mu\nu}$, 
%where $\theta^{\mu}$ is a local coordinate on $\calX$.
%\noindent
%{\bf Definition}

A (classical) P-structure on $\Map(\calX, \calM)$ is defined as follows:
\begin{definition} 
For a graded symplectic form $\omega$ on $\calM$, 
a graded symplectic form $\bomega$ on $\Map(\calX, \calM)$
is defined as $\bomega := \mu_* \ev^* \omega$.
%linear operator $\hat{\Delta}$ on $\Map(\calX, \calM)$ as
%$\hat{\Delta} \mu_* \ev^* F = \mu_* \ev^* \Delta F$,
%where $F \in C^{\infty}(\calM)$.
\end{definition}
We can confirm that ${\bomega}$ satisfies the definition of 
a graded symplectic form because $\mu_* \ev^*$ preserves
nondegeneracy and closedness.
Thus $\bomega$ is a P-structure on $\Map(\calX, \calM)$
and induces a graded Poisson bracket $\sbv{-}{-}$ on $\Map(\calX, \calM)$.
Since $|\mu_* \ev^*|=-4$, $|\bomega| = -1$ and 
$\sbv{-}{-}$ on $\Map(\calX, \calM)$ has degree $1$ and an odd
Poisson bracket.

%If a measure $\brho$ on $\Map(\calX, \calM)$ is assumed,
%we can construct an odd Laplace operator $\hat{\Delta}$
%on $\Map(\calX,  \calM)$ by the similar method as in 
%section 2. It is defined as
%\begin{eqnarray}
%\hat{\Delta} F := \frac{(-1)^{|F|}}{2} {\rm div}_{\brho} X_F,
%\nonumber
%\end{eqnarray}
%where $F \in C^{\infty}(\Map(\calX, \calM))$.

Next we define a Q-structure $S$ on $\Map(\calX,  \calM)$.
$S$ is called a {\it BV action} and
consists of two parts $S = S_0 + S_1$.
$S_0$ is constructed as follows: 
%
%
%
%Let 
%$\sbv{\cdot}{\cdot}$ be
%a nondegenerate graded Poisson bracket 
%induced from a P-structure on $\calM$. 
%This defines t
Let $\omega$ be the odd symplectic form 
%derived from a P-structure 
on $\calM$.
%
%(= D \bphi \wedge D \bb_1 -D \ba_1 \wedge D \ba_1)$ 
%($D=\theta^{\mu} \partial_{\mu}$) 
We take a fundamental form $\vartheta$ such that 
$\omega= - d \vartheta$ and
%$\Xi = {\rm ev}^* \vartheta$
define $S_0 := \iota_{\hat{D}} \mu_* {\rm ev}^* \vartheta$.
$|S_0|=0$ because $\mu_* {\rm ev}^*$ has degree $-4$.
$S_1$ is constructed as follows: 
We take a Q-structure $\Theta$ on $\calM$ and 
%A homological vector field 
define $S_1 := \mu_* \ev^* \Theta$.
$S_1$ also has degree $0$.
% because of $\mu_* {\rm ev}^*$.
%$|\Theta|=n \Longleftrightarrow |S|=0$.

We can prove that
$S$ is a Q-structure on $\Map(\calX,  \calM)$, 
since 
\begin{eqnarray}
\sbv{\Theta}{\Theta} =0,
%\Delta (e^{\frac{i}{\hbar} \Theta}) =0
\Longleftrightarrow 
%\hat{\Delta} (e^{\frac{i}{\hbar} S}) =0,
\sbv{S}{S} =0
\label{classicalmaster}
\end{eqnarray}
from the definition of $S_0$ and $S_1$.
% from assumption that $\Theta$ is a Q-structure on $\calM$.
%Since $S$ satisfies $\Delta (e^{\frac{i}{\hbar} S}) =0$, 
%Note that $S_0$ is defined locally, but 
%(\ref{classicalmaster}) can be confirmed globally.

A quantum version is 
\begin{eqnarray}
\Delta (e^{\frac{i}{\hbar} \Theta}) =0
\Longleftrightarrow 
\hat{\Delta} (e^{\frac{i}{\hbar} S}) =0,
\label{classicalmaster2}
\end{eqnarray}
where $\hat{\Delta}$ is an odd Laplace operator on $\Map(\calX, \calM)$.
The infinitesimal form of the right hand side 
in (\ref{classicalmaster2}) is 
$\sbv{S}{S} - 2 i \hbar \hat{\Delta} S =0$,
which is called a {\it quantum master equation}.
\footnote{Discussion for an odd Laplace operator is too naive. 
In general, the quantum master equation has an obstruction 
expressed by the modular class \cite{Lyakhovich:2004kr}.
We must regularize an odd Laplace operator
and a quantum BV action.
}

%We summarize 
%Then t
The following theorem has been confirmed
\cite{Alexandrov:1995kv}:
%\hfil\break
%\noindent
%{\bf Theorem}: 
\begin{thm}
If $\calX$ is a dg manifold and 
$\calM$ is a QP-manifold,
%with $\calM|_0=M$,
the graded manifold $\Map(\calX, \calM)$
has a QP-structure.
%induced from the QP-structure on $\calM$.
\end{thm}

%In $\hbar \longrightarrow 0$ (classical) limit,
%(\ref{classicalmaster}) reduces to
%\begin{eqnarray*}
%\sbv{\Theta}{\Theta} =0
%\Longleftrightarrow \sbv{S}{S} =0.
%\label{}
%\end{eqnarray*}
%The right hand side in this equation is 
%called a {\it classical master equation}.

%We define a quadruple ($\calX, \calM, \hat{\Delta}, S$)
%as {\it a topological field theory},
%induced from 
%constructed in the theorem from the above construction
%induced from a QP-structure on $\calM$.
%
%\noindent
%{\bf Definition}:
\begin{definition}
A {\it topological field theory} in $4$ dimensions
is a triple ($\calX, \calM, S$), 
where $\calX$ is a dg manifold with
$\dim \calX|_0 = 4$, 
$\calM$ is 
a QP-manifold
with the degree $3$,
%(or a $\Delta$-cohomology) 
%with the degree $n$, 
%$\hat{\Delta}$ is an odd Laplace operator 
%with the total degree $-3$
and $S$ is a BV action with the total degree $0$.
% such that $\sbv{S}{S} =0$.
%$\hat{\Delta} (e^{\frac{i}{\hbar} S}) =0$.
\end{definition}

%on a $\bZ_n$ graded manifold $\Map(\calX, \calM)$.
%Let $\calX, \calM$ be a graded manifold and
%$\dim \calX|_0 =n$.
%, where $\calX|_0$ is a degree $0$ part of $\calX$.

%We define a P-structure on $\Map(\calX \calM)$.
In order to interpret this theory
as a `physical' topological field theory,
we must take $\calX= 
%$ a special graded manifold, $
T[1]X$.
%and $\calM$ is a $\bZ_n$ graded manifold.
%, so far.
Then we can confirm that 
a QP-structure on $\Map(\calX, \calM)$ is 
equivalent to the AKSZ formulation of a topological field theory
\cite{Cattaneo:2001ys}\cite{Ikeda:2006wd}.
We set $\calX = T[1]X$ from now.

%Physically we can prove that
%a QP-manifold 
%(or a $\Delta$-cohomology) 
%with the degree $n$
%on a $\bZ_n$ graded manifold $\Map(\calX, \calM)$
%is equivalent the AKSZ-BV formalism of a topological field theory

In `physics', a quantum field theory is constructed by quantizing a
classical field theory.
%That is, first 
First we consider a Q-structure 
$\sbv{\cdot}{\cdot}$ and 
a classical P-structure $S$ such that
\begin{eqnarray*}
\sbv{S}{S} =0.
\end{eqnarray*}
Next we define 
%a measure $\brho$ on $\calM$ and derive 
a quantum P-structure $\hat{\Delta}$ 
%by the method in section 2, 
and confirm that
%find 
%a QP-manifold $\tilde{{\calM}}$ which includes $\calM$,
%a P-structure 
%$\tilde{\Delta}$
%and a Q-structure $\tilde{S}$
\begin{eqnarray*}
\tilde{\Delta} (e^{\frac{i}{\hbar} S}) =0.
\end{eqnarray*}
Finally we calculate a partition function 
\begin{eqnarray*}
Z = \int_{\calL} e^{\frac{i}{\hbar} S},
\end{eqnarray*}
on a Lagrangian submanifold 
$\calL \subset \Map(\calX, {\calM})$.
Quantization is not discussed in this paper.

%The quantization procedure is following.
%(\ref{divergence})
%(\ref{divergence2})

%%%%%%%%%%%%%%%%%%%%%%%%%%%%%%%%%%%%%%%%%%%%%%%%%%%%%%%%%%%%%%%%%%%%%
\subsection{Local Coordinate Expression and Examples}
\noindent
A general theory in the previous subsection
is applied to the local coordinate expression 
in section 3.1 and 
a known topological field theory in $4$ dimensions
is obtained as a special case and 
a new nontrivial topological field theory is constructed.
Let us take a manifold $X$ in $4$ dimensions and
a manifold $M$ in $d$ dimensions.
Let $E[1]$ is a graded vector bundle on $M$.
We take $\calX = T[1]X$ and $\calM = T^*[3]E[1]$.

Let $(\sigma^{\mu}, \theta^{\mu})$ be a local coordinate
on $T[1]X$. $\sigma^{\mu}$ is a local coordinate on 
the base manifold $X$ and 
$\theta^{\mu}$ is one on the fiber of $T[1]X$, respectively.
Let $\bbx^i$ be a smooth map $\bbx^i: X \longrightarrow M$ and 
$\bbxi_i$ be a section of $T^*[1]X \otimes \bbx^*(T^*[3] M)$, 
$\bbq^a$ be a section of $T^*[1]X \otimes \bbx^*(E[1])$ and
$\bbp_a$ be a section of $T^*[1]X \otimes \bbx^*(T^*[3]E_{\bbx}[1])$.
These are called {\it superfields}.
The exterior derivative $d$ is taken 
as a differential $D$ on $X$.
From $d$, a differential 
$\bbd = \theta^{\mu} \frac{\partial}{\partial \sigma^{\mu}}$
on $\calX$ is induced.

%In a local coordinate, 
Then a BV action $S$ has the following expression:
\begin{eqnarray*}
S&=&S_0 + S_1,\\
S_0&=&\int_{\calX} \mu \ (\bbxi_i \bbd \bbx^i
- \bbp_a \bbd \bbq^a),\\
S_1&=&\int_{\calX} \mu \ (f{}_1{}^i{}_{a} (\bbx) \bbxi_i \bbq^a 
+ \frac{1}{2} f_2{}^{ab}(\bbx) \bbp_a \bbp_b
+ \frac{1}{2} f_3{}^a{}_{bc}(\bbx) \bbp_a \bbq^b \bbq^c
+ \frac{1}{4!} f_4{}_{abcd}(\bbx) \bbq^a \bbq^b \bbq^c \bbq^d).
%\label{4DBVaction}
\end{eqnarray*}
%
%\subsection{Examples}
%\noindent
\medskip\\
\noindent
\textbf{Nonabelian BF theory}.
Let $\Theta$ be a Q-structure (\ref{liealgebra})
for a Lie algebra $\mathfrak{g}$.
$\bbxi_i \bbd \bbx^i=0$ since $M = \{pt\}$.
If we define a curvature
$\bbF^a = \bbd \bbq^a - \frac{1}{2} f{}^a{}_{bc} \bbq^b \bbq^c$,
a Q-structure is
\begin{eqnarray*}
&& S= \int_{\calX} \mu \ (- \bbp_a \bbF^a),
\end{eqnarray*}
which is equivalent to a BV formalism for
a nonabelian BF theory in $4$ dimensions.
%A local coordinate on $\Map(\calX, \calM)$ is called a superfield.
\medskip\\
\noindent
\textbf{Topological Yang-Mills Theory}.
We take a nondegenerate Killing form $(\cdot,\cdot)_{K}$
for a Lie algebra $\mathfrak{g}$ and consider the Q-structure 
(\ref{killingQ}).
A topological field theory constructed from (\ref{killingQ})
is 
\begin{eqnarray*}
&& S= \int_{\calX} \mu \ (- \bbp_a \bbF^a
+ k^{ab} \bbp_a \bbp_b).
\end{eqnarray*}
This is equivalent to a topological Yang-Mills theory,
\begin{eqnarray*}
&& S= - \frac{1}{4} \int_{\calX} \mu \ k_{ab} \bbF^a \bbF^b,
\end{eqnarray*}
if we delete $\bbp_a$ by the equations of motion.
\medskip\\
\noindent
\textbf{Nonassociative  BF Theory}.
Let us take a non Lie algebra $(\mathfrak{g},[\cdot,\cdot],(\cdot,\cdot)_{K})$
in section 4.3.
If we take
$M = \{pt\}$ and $\calM = \mathfrak{g^*}[2] \oplus \mathfrak{g}[1]$, 
$(\mathfrak{g},[\cdot,\cdot],(\cdot,\cdot)_{K})$ leads
a QP-structure with degree $3$.
In the canonical basis, it is expressed as
\begin{eqnarray*}
&& f{}_1{}^i{}_{a} (\bx) = 0, \qquad 
f_2{}^{ab}(\bx) = K^{ab}, \\
&& f_3{}^a{}_{bc}(\bx) = f{}^a{}_{bc}, \qquad 
f_4{}_{abcd}(\bx) = K^{-1}_{ae} f{}^e{}_{f[b} f{}^f{}_{cd]},
\end{eqnarray*}
where 
$K^{ab} = (p_a, p_b)$ is nondegenerate and 
$[p_a, p_b] = f{}^c{}_{ab} p_c$
is a nonassociative bracket 
and does not satisfy the Jacobi identity.
The AKSZ construction derives a new nontrivial topological field theory
in $4$ dimensions.
A BV action $S$ has the following expression:
\begin{eqnarray*}
S&=&
\int_{\calX} \mu \ (
- \bbp_a \bbd \bbq^a
+ \frac{1}{2} K{}^{ab} \bbp_a \bbp_b
+ \frac{1}{2} f{}^a{}_{bc} \bbp_a \bbq^b \bbq^c
+ \frac{1}{4!} 
K^{-1}_{ae} f{}^e{}_{f[b} f{}^f{}_{cd]}
\bbq^a \bbq^b \bbq^c \bbq^d)\\
&=& - \frac{1}{4} \int_{\calX} \mu \ (K_{ab} \bbF^a \bbF^b+
\frac{1}{3!} K^{-1}_{ae} f{}^e{}_{f[b} f{}^f{}_{cd]}
\bbq^a \bbq^b \bbq^c \bbq^d).
%\label{4DBVaction}
\end{eqnarray*}
It is easily confirmed that $\sbv{S}{S}=0$. 
\medskip\\
\noindent
\textbf{Topological $3$-brane on $Spin(7)$-structure}.
Let $(M, \Omega)$ be an $8$-dimensional $Spin(7)$-manifold.
Here $\Omega$ is a $Spin(7)$ $4$-form, 
which satisfies $d \Omega=0$ and 
the selfdual condition $\Omega=*\Omega$.
A $Spin(7)$ structure is defined as the subgroup of $GL(8)$ to 
preserve $\Omega$.
%\cite{Joyce}.
The Q-structure on $(TM,\Omega)$ is given by
\begin{eqnarray}
\Theta = \bxi_i \bq^i 
+ \frac{1}{4!} \Omega{}_{ijkl}(\bx) \bq^i \bq^j \bq^k \bq^l.
\label{multisymQ}
\end{eqnarray}
The BV action $S$ for (\ref{multisymQ}) defines the 
same theory as 
the topological $3$-brane analyzed in \cite{Bonelli:2005ti}.

%%%%%%%%%%%%%%%%%%%%%%%%%%%%%%%%%%%%%%%%%%%%%%%%%%%%%%%%%%%%%%%%%%%%%
%%%%%%%%%%%%%%%%%%%%%%%%%%%%%%   SEC  6    %%%%%%%%%%%%%%%%%%%%%%%%%%
%%%%%%%%%%%%%%%%%%%%%%%%%%%%%%%%%%%%%%%%%%%%%%%%%%%%%%%%%%%%%%%%%%%%%
\section{Conclusions and Discussion}
\noindent
We have defined a BV algebra and a QP-structure 
of degree $3$.
A QP-structure of degree $3$ has been constructed explicitly 
and a Lie algebroid u.t.h.~has been defined as 
its algebraic and geometric structure.
A general theory of the AKSZ construction of a 
topological field theory 
%in four dimensions 
has been expressed and
a new topological field theory in four 
dimensions has been constructed
from a QP-structure.
% with the degree $4$.

Quantization of this theory and
analysis of a Lie algebroid u.t.h.~will 
shed light on a super Poisson geometry and a quantum field theory.
They are future problems.

%%%%%%%%%%%%%%%%%%%%%%%%%%%%%%%%%%%%%%%%%%%%%%%%%%%%%%%%%%%%%%%%%%%%%
%%%%%%%%%%%%%%%%%%%%%%%%%%%%%%   APPENDIX    %%%%%%%%%%%%%%%%%%%%%%%%
%%%%%%%%%%%%%%%%%%%%%%%%%%%%%%%%%%%%%%%%%%%%%%%%%%%%%%%%%%%%%%%%%%%%%
%\section*{Appendix}
%\noindent
%

%%%%%%%%%%%%%%%%%%%%%%%%%%%%%%%%%%%%%%%%%%%%%%%%%%%%%%%%%%%%%%%%%%%%%
%%%%%%%%%%%%%%%%%%%%%%%%%%%%   ACKNOWLEGE    %%%%%%%%%%%%%%%%%%%%%%%%
%%%%%%%%%%%%%%%%%%%%%%%%%%%%%%%%%%%%%%%%%%%%%%%%%%%%%%%%%%%%%%%%%%%%%
\section*{Acknowledgements}

The authors would like to thank Klaus Bering, Maxim Grigoriev, 
Camille Laurent-Gengoux, Yvette Kosmann-Schwarzbach, Kirill Mackenzie, 
Dmitry Roytenberg, Alexei Sharapov, Thomas Strobl and Theodore Voronov
for their comments and discussion.
The author (N.I.)  would like to thank Maskawa Institute for Science and 
Culture, Kyoto Sangyo University for hospitality.
We would like to thank to referees for their useful advice.

\newcommand{\bibit}{\sl}

%%%%%%%%%%%%%%%%%%%%%%%%%%%%%%%%%%%%%%%%%%%%%%%%%%%%%%%%%%%%%%%%%%%%%
%%%%%%%%%%%%%%%%%%%%%%%%%%%%%%%  Refs. %%%%%%%%%%%%%%%%%%%%%%%%%%%%%%
%%%%%%%%%%%%%%%%%%%%%%%%%%%%%%%%%%%%%%%%%%%%%%%%%%%%%%%%%%%%%%%%%%%%%
%\newpage
%NEW MACRO FOR BIBLIOGRAPHY

%\section*{References}
%\noindent

\vfill\eject
\end{document}